# Entropy associated with conformation and density fluctuations in biomolecular solutions

Fumio Hirata[1]

(hirata@ims.ac.jp)

*National Institutes of Natural Sciences, Institute for Molecular Science, Myodaiji, Okazaki, Aichi 444-8585, Japan.*

[1]Professor Emeritus

**Abstract**

Microscopic formula to describe the entropy of biomolecular solutions are derived based on the Gibbs formula of entropy, and the generalized Langevin theory combined with the RISM/3D-RISM theory. Two formula are derived: one is concerned with the conformational fluctuation of a biomolecule, and the other with the density fluctuation of solvent around a solute. The formula derived for the entropy associated with the conformational fluctuation is

$$S_{conf} = \frac{1}{2} k_B \log\left(\frac{(2\pi)^{3N}}{A}\right) + \frac{3N}{2} k_B$$

where $N$ is the number of atoms in the solute, and $A$ is the determinant of the inverse of the variance-covariance matrix of conformational fluctuation.

The formula for the entropy of solvent at a pair of positions around a solute is also derived to be,

$$s_{solv}(\mathbf{r},\mathbf{r}') = \frac{1}{2} k_B \log\left(\frac{(2\pi)^{n}}{B}\right) + \frac{3n}{2} k_B$$

where $n$ is the number of atoms in a solvent molecule, and $B$ is essentially the determinant of the matrix of the density-pair-correlation functions. The entropy at a local position $\mathbf{r}$ may be obtained by integrating the expression by $\mathbf{r}'$ over the entire volume of the system.

The feasibility of the calculation to find the entropies is discussed.

## I. Introduction

Entropy, a thermodynamic variable, has been attracting increasing attention from various fields of bioscience and technology. [1-8] One of those phenomena in which entropy plays a crucial role is the conformational stability of a biomolecule in solution. It is well documented that the denatured state of protein is dominated by entropy.[1,2,9-11] Another phenomenon, that is concerned with entropy, is the binding affinity of a ligand molecule at the active site of a protein molecule. [3,5~8] In that case, the binding affinity is determined not only by the interaction between side-chains in the active-site and the ligand molecule, but also by gain or loss of the entropy related to the protein and solvent molecules. Although it is of great importance to evaluate the quantity by means of computational science, actual calculation is not so straightforward. It is because the quantity reflects *fluctuation* of mechanical as well as thermodynamic variables, which is a stochastic process.

The entropy relevant to biomolecules in solution is classified into two categories, depending on whether it is related to the interaction among molecules or not. [12, 13] Those which are not relevant to the intermolecular interactions include the *ideal* and *mixing* entropies. The ideal entropy is originated from the kinetic motion of individual molecules in the system, which can be calculated readily from the classical mechanics of a single molecule as a function of single-molecular properties such as the mass, moment of inertia, and force constant. [14] The mixing entropy is just a function of the concentration of the system, the calculation of which is also trivial as $-\sum_i k_B \log x_i$, where $x_i$ denotes the concentration (*mol fraction*) of species *i* in the solution.[12, 13] So, it is the entropy related to the interaction among atoms and molecules that has not been fully explored yet. Let us refer to the entropy as *excess entropy*.

There are two types of the excess entropy that is intimately related to the biomolecular processes in solution, such as protein folding and the binding of a substrate molecule to an enzyme. The first one of those is the structural or conformational entropy which is intimately related to the temporal fluctuation of the structure of biomolecules in solution. Since the fluctuation from an equilibrium structure takes place against the free energy slope in a stochastic process, it is the main concern to find the probability distribution of a conformational fluctuation. The other type of the excess entropy is concerned with the density fluctuation of solvent around a biomolecule. For examples, the density of solvent molecules inside a cavity of an enzyme is in temporal fluctuation, conjugated with the conformational fluctuation of side-chains conforming the cavity. Such fluctuation of solvent density around a biomolecule will make contribution to the entropy change upon binding of a substrate to the active site of an enzyme.

The first serious attempt to realize the conformational entropy of a biomolecule in the statistical-mechanical framework has been made by Go, Go, and Scheraga. [15-17] They have assumed the probability distribution of the atomic displacement around the dihedral angles as *Gaussian*, and calculated the conformational entropy of protein using the Boltzmann formula. Unfortunately, the entropy formula they have derived was not so realistic, since the force-constant matrix they have used is concerned with a *protein in vacuum*. There are two problems to be pointed out concerning *protein in vacuum* to evaluate the conformational entropy of protein. Firstly, the equilibrium structure of protein *in vacuum* is entirely different from that in aqueous solution. Secondly, the conformational fluctuation, or the variance-covariance matrix, may also be largely different from the real system. It should be noted also that the Gaussian character of the conformational distribution has its origin in the truncation of the interaction potential energy at the second order of the Taylor expansion with respect to the displacement of atoms from the equilibrium structure.

Kusick and Karplus put the calculation of the conformational entropy of protein on the foundation of the molecular-dynamics simulation. [18] They have analyzed the variance-covariance matrix of a protein over the MD trajectory to find the conformational entropy, which reflects the structural fluctuations of protein. The entropy so obtained certainly includes the contribution from multiple conformations within a limit of the length of the trajectory sampled by the simulation.

Recently, Chong and Ham have proposed a new computational method to estimate the conformational entropy of protein in water based on the RISM/3D-RISM theory combined with the MD simulation. [1,2] They have carried out the cumulant-expansion of the free energy of solute, the structure of which is fluctuating with certain probability. When the probability distribution is Gaussian, the expansion can be truncated by the second order of the fluctuation to produce an expression for the conformational entropy of a protein as

$$S_{conf} = \frac{k_B}{2} \left\langle \left( F - \langle F \rangle \right)^2 \right\rangle \quad (1)$$

In Eq. (1), $F$ is the free energy of solute, defined by $F \equiv E + \Delta\mu$, where $E$ and $\Delta\mu$ denote the interatomic potential-energy of protein and the solvation free energy, respectively, and $<F>$ is the average of $F$ over the fluctuation. The authors have calculated $E$ and $\Delta\mu$ along the trajectory of the MD simulation of protein in aqueous solution. In order to calculate $\Delta\mu$, they have employed the RISM/3D-RISM theory. An important observation made by the authors is that the probability distribution of the free-energy fluctuation is *Gaussian*, that rationalizes the use of Eq. (1) to calculate entropy. The observation is in harmony with the theoretical characterization of the conformational fluctuation of protein in solution, made by Kim and Hirata.[19] Based on the generalized

Langevin equations, derived for the conformational fluctuation of protein, Kim and Hirata have identified that the fluctuation is taking place on the free energy surface which is quadratic with respect to the displacement of atoms from their equilibrium structure. The identification is equivalent to assert the conformational fluctuation being *Gaussian*.

In the present paper, we develop a theory to calculate the entropy of a system consisting of a biomolecule and solvent, based on the Kim-Hirata theory. Here, we are interested in two type of entropies which are associated with the conformational fluctuation of solute and the density fluctuation of solvent around the solute.

In order to formulate the entropies, we employ the Gibbs entropy formula. [20] Suppose we have an event $x$ which is fluctuating around its average value $\langle x \rangle$. Then, the Gibbs entropy of the event may be written as

$$S = -k_B \int w(\Delta x) \log w(\Delta x) dx, \qquad (2)$$

where $\Delta x (= x - \langle x \rangle)$ represents the fluctuation of the event $x$ from its average value, $\langle x \rangle$, and $w(\Delta x)$ denotes the probability dsitribution of the event to take a value $\Delta x$. So, it is crucial in our case to find the *probability distribution* of the two types of fluctuation, the conformational fluctuation of a solute and the density fluctuation of solvent around the solute.

The *excess* entropy associated with a protein in solution is originated from two types of fluctuations: the conformational fluctuation of protein in solution and the density fluctuation of solvent around the solute.

$$S_{excess} = S_{conf} + S_{solv} \qquad (3)$$

where $S_{conf}$ and $S_{solv}$ denote the entropies associated with the conformational fluctuation and the density fluctuation, respectively. It should be noted that $S_{conf}$ includes a contribution from solvent around solute. However, the contribution should not be confused with $S_{solv}$. $S_{solv}$ is the entropy associated with the density fluctuation of solvent around solute with a *fixed* conformation. On the other hand, the contribution from solvent included in $S_{conf}$ is concerned with the density fluctuation of solvent conjugated with the conformational fluctuation of the solute. So, it is a cross term of the conformational fluctuation of solute and density fluctuation of solvent.

It is the purpose of the paper to derive formula to calculate the entropies based on the Kim-Hirata theory of the conformational fluctuation of a biomolecule and density fluctuation of solvent around the solute in solution.[19]

**II. Theory**
**II-1. Brief review of Kim-Hirata theory**

*Generalized Langevin Theory:* The generalized Langevin theory for a biomolecule in water, developed by Kim and Hirata, is briefly reviewed for the purpose of clarifying physical meanings of the variables employed in the derivation of entropy formula. [19, 21, 22]

The generalized Langevin equations (GLE) describes the time evolution of dynamic variables $\mathbf{A}(t)$ in the phase space, which is governed by the Liouville operator $iL$.

$$\frac{d\mathbf{A}(t)}{dt} = iL\mathbf{A}(t) \qquad (4)$$

The dynamic variables represent a set of few mechanical variables which are essential to describe the physics of interest. It is the whole idea of GLE to *project* all variables in the phase space onto the dynamic variables by means of the projection-operator method. [21,22]

In the Kim-Hirata theory, the following set of variables in the phase space was chosen for the dynamical variables.

$$\mathbf{A}(t) = \begin{pmatrix} \Delta\mathbf{R}_\alpha(t) \\ \mathbf{P}_\alpha(t) \\ \delta\rho_a(\mathbf{r},t) \\ \mathbf{J}_a(\mathbf{r},t) \end{pmatrix}, \qquad (5)$$

where $\Delta\mathbf{R}_\alpha(t)$ and $\mathbf{P}_\alpha(t)$ are the displacement or fluctuation of an atom $\alpha$ from its equilibrium position in a biomolecule, and its conjugated momentum, defined respectively by

$$\Delta\mathbf{R}_\alpha(t) \equiv \mathbf{R}_\alpha(t) - \langle\mathbf{R}_\alpha\rangle \quad \text{and} \quad \mathbf{P}_\alpha(t) \equiv M_\alpha \frac{d\Delta\mathbf{R}_\alpha}{dt}. \qquad (6)$$

where $\langle\mathbf{R}_\alpha\rangle$ and $M_\alpha$ denote the equilibrium position of the atom $\alpha$ and its mass, respectively. $\delta\rho_a(\mathbf{r},t)$ is the density fluctuation of atom $a$ of solvent molecules at position **r**, defined by,

$$\delta\rho_a(\mathbf{r},t) = \sum_i \delta(\mathbf{r}-\mathbf{r}_i^a(t)) - \langle \rho_a \rangle, \tag{7}$$

where $\langle \rho_a \rangle$ defines the average density of solvent atom *a*. $\mathbf{J}_a(t)$ is the momentum density, or the current, of solvent atom *a*, defined by

$$\mathbf{J}_a(\mathbf{r},t) \equiv \sum_i \mathbf{p}_i^a \delta(\mathbf{r}-\mathbf{r}_i^a). \tag{8}$$

Applying the projection operator method to the Liouville equation, Eq. (4), Kim and Hirata have derived two GLEs, one for the biomolecule, and the other for solvent.

***GLE of a biomolecule in solution:*** The conformational fluctuation of a biomolecule in solution is described by

$$M_\alpha \frac{d^2 \Delta \mathbf{R}_\alpha(t)}{dt^2} = -\sum_\beta A_{\alpha\beta} \Delta \mathbf{R}_\beta(t) - \int_0^t ds \sum \Gamma_{\alpha\beta}(t-s) \cdot \frac{d\Delta \mathbf{R}_\alpha(s)}{ds} + W_\alpha(t), \tag{9}$$

where $A_{\alpha\beta}$ is the $(\alpha,\beta)$ element of the inverese of the variance-covariance matrix defined by,

$$\mathbf{L} \equiv \langle \Delta \mathbf{R} \Delta \mathbf{R} \rangle. \tag{10}$$

Let us express $A_{\alpha\beta}$ by,

$$A_{\alpha\beta} = k_B T \left( \mathbf{L}^{-1} \right)_{\alpha\beta} \tag{11}$$

The first term in the right-hand side of Eq. (9) is concerned with the restoring force which is proportional to the displacement of atoms from an equilibrium structure. Second and third terms are the frictional and random forces, which are related to each other by the *fluctuation-dissipation* theorem.[23] The physics of the equation is clear: the conformational fluctuation of atom $\Delta \mathbf{R}_\alpha(t)$ is induced by the thermal motion, or the random force $W_\alpha(t)$, and it relaxes to the equilibrium structure $\langle \mathbf{R}_\alpha \rangle$ due to the restoring force, or the first term.

If one ignores the second and third terms, Eq. (9) becomes identical to the equation describing a harmonic oscillator as,

$$M_\alpha \frac{d^2 \Delta \mathbf{R}_\alpha(t)}{dt^2} = -\sum_\beta A_{\alpha\beta} \Delta \mathbf{R}_\beta(t), \tag{12}$$

where the factor $A_{\alpha\beta}$ plays a role of the *force constant* or *Hessian*. An important *ansatz* made by the authors is to identify $A_{\alpha\beta}$ as the second derivative of the free energy surface of a biomolecule, consisting of the interatomic potential energy ($U$) and the solvation free energy ($\Delta\mu$), with respect to the atomic coordinates of the biomolecule, that is,

$$A_{\alpha\beta} = \frac{\partial^2 F}{\partial \Delta \mathbf{R}_\alpha \partial \Delta \mathbf{R}_\beta} \tag{13}$$

where

$$F(\{\Delta \mathbf{R}\}) = U(\{\Delta \mathbf{R}\}) + \Delta\mu(\{\Delta \mathbf{R}\}). \tag{14}$$

In the equation, $\{\Delta \mathbf{R}\}$ represents a set of atomic displacements due to the conformational fluctuation. The identification implies that the free energy surface is quadrature with respect to the displacement of atoms, and that the the probability distribution $w_{conf}$ of finding the conformational fluctuation of biomolecule in $\{\Delta \mathbf{R}\}$ is *"Gaussian",* or

$$w_{conf}(\{\Delta \mathbf{R}\}) = \sqrt{\frac{A}{(2\pi)^{3N}}} \exp\left[-\frac{1}{2}\sum_\alpha \sum_\beta A_{\alpha\beta} \Delta \mathbf{R}_\alpha \Delta \mathbf{R}_\beta\right] \tag{15}$$

where $A$ is the determinant of the matrix $\{A_{\alpha\beta}\}$.

***GLE for the density fluctuation of solvent around a biomolecule:*** The temporal density fluctuation of atoms of water molecules at position **r** around a biomolecule is described by

$$\frac{\partial^2 \delta\rho_a(\mathbf{r},t)}{\partial t^2} = -\sum_{b,c} \int \kappa_{ab}^{solv}(\mathbf{r},\mathbf{r}') \delta\rho_c(\mathbf{r}',t) d\mathbf{r}'$$
$$+ \left(\frac{1}{N}\sum_{b,c} \int d\mathbf{r}' \int d\mathbf{r}'' J_{ab}(\mathbf{r},\mathbf{r}') \int_0^t ds M_{bc}(\mathbf{r}';t-s) \cdot \frac{d\delta\rho_c(\mathbf{r}'',s)}{dt} ds\right) - \Xi_a(\mathbf{r},t)$$

$$\tag{16}$$

where $\kappa_{ab}^{solv}(\mathbf{r},\mathbf{r}')$ is defined by,

$$\kappa_{ab}^{solv}(\mathbf{r},\mathbf{r}') \equiv -\sum_c \int d\mathbf{r}'' J_{ac}(\mathbf{r},\mathbf{r}'') \chi_{cb}^{(2)}(\mathbf{r}'',\mathbf{r}')^{-1} \qquad (17)$$

In the equation, $J_{ac}$ is the kinetic contribution of molecules to the density fluctuation, expressed by,

$$J_{ac}(\mathbf{r},\mathbf{r}') = \sum_i \langle \mathbf{v}_i^a \cdot \mathbf{v}_i^b \rangle \omega_{ab}(|\mathbf{r}-\mathbf{r}'|), \qquad (18)$$

where $\mathbf{v}_i^a$ and $\mathbf{v}_i^b$ denote the velocity of atoms $a$ and $b$ in the same molecule $i$, and $\omega_{ab}$ is the *intra*molecular correlation-function defined by

$$\omega_{ab}(|\mathbf{r}-\mathbf{r}'|) = \rho \delta_{ab} \delta(|\mathbf{r}-\mathbf{r}'|) + \rho(1-\delta_{ab}) y_{ab}(|\mathbf{r}-\mathbf{r}'|)$$
$$y_{ab}(|\mathbf{r}-\mathbf{r}'|) = \frac{1}{4\pi l_{ab}^2} \delta(|\mathbf{r}-\mathbf{r}'|-l_{ab}) \qquad (19)$$

where $\delta_{ab}$ and $\delta(|\mathbf{r}-\mathbf{r}'|)$ denote the Kronecker and Dirac delta-functions, respectively, $l_{ab}$ is the distance, or "bond", between atoms $a$ and $b$ in a same molecule. So, the structure of a solvent molecule is defined by giving $l_{ab}$ for all pairs of atoms in the molecule. $\chi_{ab}^{(2)}$ is the density pair correlation function defined by,

$$\chi_{ab}^{(2)}(\mathbf{r},\mathbf{r}') \equiv \langle \delta\rho_a(\mathbf{r}) \delta\rho_b(\mathbf{r}') \rangle \qquad (20)$$

The first term in the right-hand side in Eq. (16) represents the restoring force to recover the equilibrium density of solvent atom $a$ at the position $\mathbf{r}$, which is proportional to the density fluctuation $\delta\rho_c(\mathbf{r}',t)$ at the position $\mathbf{r}'$. The second and third terms are the frictional and random forces, which are related each other by the fluctuation-dissipation theorem.

Since we are interested in the entropy associated with the *temporal* density-fluctuation at a fixed position, we just focus on the first term of the right-hand-side in Eq. (16). That leads,

$$\frac{\partial^2 \delta\rho_a(\mathbf{r},t)}{\partial t^2} = -\sum_b \int \kappa_{ab}^{solv}(\mathbf{r},\mathbf{r}')\delta\rho_b(\mathbf{r}',t)d\mathbf{r}' \qquad (21)$$

The equation can be interpreted as follows. The density fluctuation of atom *a*, induced at position **r**, is restored to the equilibrium density by the force which is exerted from the density fluctuation of atom *b* located at **r'**, and the force constant of the restoring force is $\kappa_{ab}^{solv}(\mathbf{r},\mathbf{r}')$.

In order to complete the mathematical isomorphism with a harmonic oscillator, we define the free energy of solvent by,

$$F_{solv} \equiv \int\int \mu_{solv}(\mathbf{r},\mathbf{r}')d\mathbf{r}d\mathbf{r}' \qquad (22)$$

where the *local free energy* $\mu_{solv}(\mathbf{r},\mathbf{r}')$ at a pair of positions **r** and **r'** is defined by

$$\mu_{solv}(\mathbf{r},\mathbf{r}') \equiv \frac{1}{2}k_B T \sum_a \sum_b \delta\rho_a(\mathbf{r})\kappa_{ab}^{solv}(\mathbf{r},\mathbf{r}')\delta\rho_b(\mathbf{r}') \qquad (23)$$

Then, $\kappa_{ab}^{solv}(\mathbf{r},\mathbf{r}')$ is identified as the second derivative of the free energy $F_{solv}$ with respect to the density fluctuation at fixed positions **r** and **r'**,

$$\kappa_{ab}^{solv}(\mathbf{r},\mathbf{r}') = \frac{\delta^2(F_{solv}/k_B T)}{\delta\rho_a(\mathbf{r})\delta\rho_b(\mathbf{r}')} \qquad (24)$$

The identification is similar to the density-functional theory of liquids, except for the kinetic factor defined by Eq. (18). [19] The kinetic factor is included there, because the momentum density or flux also contributes to change of the free energy.

The physics of the equation can be interpreted in the analogy to the harmonic oscillator as follows. The *force* to restore the *equilibrium density* of solvent-atom *a* at position **r** is proportional to the density fluctuation of solvent-atom *b* at position **r'**, and the *force constant* $\kappa_{ab}^{solv}(\mathbf{r},\mathbf{r}')$ of the *restoring force* is the second derivative of the free energy with respect to the density fluctuation of atoms *a* and *b*.

The probability distribution of the density fluctuation of solvent atoms *a* and *b* at **r** and **r'**, having values $\delta\rho_a(\mathbf{r})$ and $\delta\rho_b(\mathbf{r}')$, respectively, is

$$w_{solv}(\{\delta\rho\};\mathbf{r},\mathbf{r}') \propto \exp\left[-\frac{\mu_{solv}(\mathbf{r},\mathbf{r}')}{k_B T}\right]$$
$$= \sqrt{\frac{B}{(2\pi)^{3n}}} \exp\left[-\frac{1}{2}\sum_a\sum_b \delta\rho_a(\mathbf{r})\kappa_{ab}^{solv}(\mathbf{r},\mathbf{r}')\delta\rho_a(\mathbf{r}')\right] \quad (25)$$

where $B$ is the normalization constant, and it is the determinant of the matrix $\{\kappa_{ab}^{solv}(\mathbf{r},\mathbf{r}')\}$. The expression is apparently a Gaussian distribution with respect to the density fluctuations at the fixed position **r** and **r**', the variance of which is $\kappa_{ab}^{solv}(\mathbf{r},\mathbf{r}')^{-1}$.

**II-2. Entropy associated with conformational fluctuation of a solute molecule**

The entropy associated with the conformational fluctuation, defined in terms of the Gibbs formula (Eq. (1)) is,

$$S_{conf} = -k_B \int w_{conf}(\{\Delta\mathbf{R}\}) \log w_{conf}(\{\Delta\mathbf{R}\}) \prod_\alpha d\Delta\mathbf{R}_\alpha \quad (26)$$

where $\Delta\mathbf{R}_\alpha (= \mathbf{R}_\alpha - \langle\mathbf{R}_\alpha\rangle)$ represents the positional fluctuation of atom $\alpha$ in protein, and $w(\{\Delta\mathbf{R}\})$ is the probability distribution of the conformational fluctuation defined by Eq. (15). Substituting Eq. (15) into Eq. (26) leads,

$$S_{conf} = -k_B \int\cdots\int \sqrt{\frac{A}{(2\pi)^{3N}}} \exp\left[-\frac{1}{2}\sum_\alpha\sum_\beta A_{\alpha\beta}\Delta\mathbf{R}_\alpha\Delta\mathbf{R}_\beta\right] \log\left[\sqrt{\frac{A}{(2\pi)^{3N}}} \exp\left[-\frac{1}{2}\sum_\alpha\sum_\beta A_{\alpha\beta}\Delta\mathbf{R}_\alpha\Delta\mathbf{R}_\beta\right]\right] \prod_\alpha d\Delta\mathbf{R}_\alpha.$$
(27)

where $A$ is the determinant of the matrix $\{A_{\alpha\beta}\}$ that is the inverse of the variance-covariance matrix, $\{L_{\alpha\beta}\} \equiv \langle\Delta\mathbf{R}_\alpha\Delta\mathbf{R}_\beta\rangle$.

The integral in Eq. (27) can be split into two terms as follows.

$$S_{conf} = -k_B \int \cdots \int \sqrt{\frac{A}{(2\pi)^{3N}}} \exp\left[-\frac{1}{2}\sum_\alpha \sum_\beta A_{\alpha\beta}\Delta\mathbf{R}_\alpha \Delta\mathbf{R}_\beta\right] \log\left\{\sqrt{\frac{A}{(2\pi)^{3N}}} \exp\left[-\frac{1}{2}\sum_\alpha \sum_\beta A_{\alpha\beta}\Delta\mathbf{R}_\alpha \Delta\mathbf{R}_\beta\right]\right\} \prod_\alpha d\Delta\mathbf{R}_\alpha$$

$$= -k_B \int \cdots \int \sqrt{\frac{A}{(2\pi)^{3N}}} \exp\left[-\frac{1}{2}\sum_\alpha \sum_\beta A_{\alpha\beta}\Delta\mathbf{R}_\alpha \Delta\mathbf{R}_\beta\right] \left\{\log\sqrt{\frac{A}{(2\pi)^{3N}}} - \frac{1}{2}\sum_\alpha \sum_\beta A_{\alpha\beta}\Delta\mathbf{R}_\alpha \Delta\mathbf{R}_\beta\right\} \prod_\alpha d\Delta\mathbf{R}_\alpha \quad (28)$$

$$= -k_B \int \cdots \int \sqrt{\frac{A}{(2\pi)^{3N}}} \exp\left[-\frac{1}{2}\sum_\alpha \sum_\beta A_{\alpha\beta}\Delta\mathbf{R}_\alpha \Delta\mathbf{R}_\beta\right] \log\sqrt{\frac{A}{(2\pi)^{3N}}} \prod_\alpha d\Delta\mathbf{R}_\alpha$$

$$+ k_B \int \cdots \int \sqrt{\frac{A}{(2\pi)^{3N}}} \exp\left[-\frac{1}{2}\sum_\alpha \sum_\beta A_{\alpha\beta}\Delta\mathbf{R}_\alpha \Delta\mathbf{R}_\beta\right] \left\{\frac{1}{2}\sum_\alpha \sum_\beta A_{\alpha\beta}\Delta\mathbf{R}_\alpha \Delta\mathbf{R}_\beta\right\} \prod_\alpha d\Delta\mathbf{R}_\alpha$$

$$= S_1 + S_2$$

The Gaussian integral of the first term can be readily carried out to give

$$S_1 = -k_B \int \cdots \int \sqrt{\frac{A}{(2\pi)^3}} \exp\left[-\frac{1}{2}\sum_\alpha \sum_\beta A_{\alpha\beta}\Delta\mathbf{R}_\alpha \Delta\mathbf{R}_\beta\right] \log\sqrt{\frac{A}{(2\pi)^{3N}}} \prod_\alpha d\mathbf{R}_\alpha$$

$$= \frac{1}{2} k_B \log\left(\frac{(2\pi)^{3N}}{A}\right) \quad (29)$$

In order to carry out the integral in the second term of Eq. (28), we introduce a parameter $x$.

$$Y = k_B \int \cdots \int \sqrt{\frac{A}{(2\pi)^{3N}}} \exp\left[-x\frac{1}{2}\sum_\alpha \sum_\beta A_{\alpha\beta}\Delta\mathbf{R}_\alpha \Delta\mathbf{R}_\beta\right] \prod_\alpha d\mathbf{R}_\alpha . \quad (30)$$

Then, the second term in Eq. (28) can be then written as

$$S_2 = -\left.\frac{dY}{dx}\right|_{x=1} \quad (31)$$

The integral in Eq. (30) can be carried out readily to give

$$Y = k_B \int \sqrt{\frac{A}{(2\pi)^{3N}}} \exp\left[-\frac{1}{2}\sum_\alpha \sum_\beta x A_{\alpha\beta}\Delta\mathbf{R}_\alpha \Delta\mathbf{R}_\beta\right] \prod_\alpha d\Delta\mathbf{R}_\alpha$$

$$= k_B \sqrt{\frac{A}{(2\pi)^{3N}}} \sqrt{\frac{(2\pi)^{3N}}{x^{3N} A}} = k_B x^{-3N/2} \quad (32)$$

Therefore,

$$S_2 = -\left.\frac{dY}{dx}\right|_{x=1} = \frac{3N}{2} k_B \quad (33)$$

We find finally for the conformational entropy of protein as

$$S_{conf} = S_1 + S_2$$
$$= \frac{1}{2}k_B \log\left(\frac{(2\pi)^{3N}}{A}\right) + \frac{3N}{2}k_B \quad (34)$$

In the equation, $A$ is the determinant of the matrix $\{A_{\alpha\beta}\}$, the element of which is the second derivative of the free energy, defined by Eq. (14), with respect to the coordinates of atom $\alpha$ and $\beta$. It is in turn the inverse of the determinant of the variance-covariance matrix as is defined by Eqs. (10) and (11). So, the expression is in accord with our intuition, that is, *greater the variance, larger the conformational entropy*. The second derivative of the free energy may be obtained from the RISM/3D-RISM theory in the route of solving the equation iteratively. [29]

**II-3. Entropy associated with solvent fluctuation around a biomolecule in solution**

We define a quantity referred to as the *local solvent entropy* $s_{solv}(\mathbf{r},\mathbf{r}')$ at a pair of positions, $\mathbf{r}$ and $\mathbf{r}'$, as

$$s_{solv}(\mathbf{r},\mathbf{r}') \equiv -k_B \int \cdots \int w_{solv}(\{\delta\rho\};\mathbf{r},\mathbf{r}') \log w_{solv}(\{\delta\rho\};\mathbf{r},\mathbf{r}') \prod_a d\delta\rho_a, \quad (35)$$

where $w_{solv}$ is defined by Eq. (25).

It should be noted that the variable of integral is $\delta\rho(\mathbf{r})$, not the coordinate $\mathbf{r}$, since $\mathbf{r}$ is fixed. The density fluctuation here is a temporal fluctuation rather than a spatial fluctuation. The entropy of sovation at position $\mathbf{r}$ around a biomolecule, and of the entire system is defined, respectively, as

$$s_{solv}(\mathbf{r}) \equiv \int \int s_{solv}(\mathbf{r},\mathbf{r}') d\mathbf{r}' \quad (36)$$

and

$$S_{solv} \equiv \int \int s_{solv}(\mathbf{r},\mathbf{r}') d\mathbf{r} d\mathbf{r}' \quad (37)$$

The local entropy is obtained by substituting the expression of Eq. (25) for $w_{solv}$ into Eq. (36), and performing the integral with respect to the density fluctuation. The integral may be carried out as follows.

$$s_{solv}(\mathbf{r},\mathbf{r}') = -k_B \int \cdots \int \sqrt{\frac{B}{(2\pi)^{3n}}} \exp\left[-\frac{1}{2}\sum_a\sum_b \kappa_{ab}\delta\rho_a\delta\rho_b\right] \log\left\{\sqrt{\frac{B}{(2\pi)^{3n}}} \exp\left[-\frac{1}{2}\sum_a\sum_b \kappa_{ab}\delta\rho_a\delta\rho_b\right]\right\} \prod_a d\delta\rho_a$$

$$= -k_B \int \cdots \int \sqrt{\frac{B}{(2\pi)^{3n}}} \exp\left[-\frac{1}{2}\sum_a\sum_b \kappa_{ab}\delta\rho_a\delta\rho_b\right] \left\{\log\sqrt{\frac{B}{(2\pi)^{3n}}} - \frac{1}{2}\sum_a\sum_b \kappa_{ab}\delta\rho_a\delta\rho_b\right\} \prod_a d\delta\rho_a$$

$$= -k_B \int \cdots \int \sqrt{\frac{B}{(2\pi)^{3n}}} \exp\left[-\frac{1}{2}\sum_a\sum_b \kappa_{ab}\delta\rho_a\delta\rho_b\right] \log\sqrt{\frac{B}{(2\pi)^{3n}}} \prod_a d\delta\rho_a$$

$$+ k_B \int \cdots \int \sqrt{\frac{B}{(2\pi)^{3n}}} \exp\left[-\frac{1}{2}\sum_a\sum_b \kappa_{ab}\delta\rho_a\delta\rho_b\right] \left\{\frac{1}{2}\sum_a\sum_b \kappa_{ab}\delta\rho_a\delta\rho_b\right\} \prod_a d\delta\rho_a$$

$$\equiv s_{solv}^{(1)}(\mathbf{r},\mathbf{r}') + s_{solv}^{(2)}(\mathbf{r},\mathbf{r}') \tag{38}$$

where $s_{solv}^{(1)}(\mathbf{r},\mathbf{r}')$ and $s_{solv}^{(2)}(\mathbf{r},\mathbf{r}')$ are defined as,

$$s_{solv}^{(1)}(\mathbf{r},\mathbf{r}') = k_B \int \cdots \int \sqrt{\frac{B}{(2\pi)^{3n}}} \exp\left[-\frac{1}{2}\sum_a\sum_b B_{ab}\delta\rho_a\delta\rho_b\right] \log\sqrt{\frac{B}{(2\pi)^{3n}}} \prod_a d\delta\rho_a$$

$$s_{solv}^{(2)}(\mathbf{r},\mathbf{r}') = k_B \int \cdots \int \sqrt{\frac{B}{(2\pi)^{3n}}} \exp\left[-\frac{1}{2}\sum_a\sum_b B_{ab}\delta\rho_a\delta\rho_b\right] \left\{\frac{1}{2}\sum_a\sum_b B_{ab}\delta\rho_a\delta\rho_b\right\} \prod_a d\delta\rho_a \tag{39}$$

The integrals are readily carried out as follows, since they are just a gaussian integral.

$$s^{(1)}(\mathbf{r},\mathbf{r}') = -k_B \int \cdots \int \prod_a d\delta\rho_a \sqrt{\frac{B}{(2\pi)^3}} \exp\left[-\frac{1}{2}\sum_a\sum_b B_{ab}\delta\rho_a\delta\rho_b\right] \log\sqrt{\frac{B}{(2\pi)^{3N}}}$$

$$= \frac{1}{2}k_B \log\left(\frac{(2\pi)^n}{B}\right) \tag{40}$$

$$s_{solv}^{(2)}(\mathbf{r},\mathbf{r}') = k_B \int \cdots \int \prod_a d\delta\rho_a \sqrt{\frac{B}{(2\pi)^{3n}}} \exp\left[-\frac{1}{2}\sum_a\sum_b B_{ab}\delta\rho_a\delta\rho_b\right] \left\{\frac{1}{2}\sum_a\sum_b B_{ab}\delta\rho_a\delta\rho_b\right\}$$

$$= \frac{3n}{2}k_B \tag{41}$$

In the equation, $B$ is the determinant of the matrix $\{\kappa_{ab}^{solv}(\mathbf{r},\mathbf{r}')\}$ at the fixed position $\mathbf{r}$ and $\mathbf{r}$', the element of which is the inverse of the site-site pair correlation function multiplied by the kinetic factor $\{J_{ac}\}$ as defined by Eq. (18). Finally, we find an expression for the local solvent entropy $s_{solv}(\mathbf{r},\mathbf{r}')$ as,

$$s_{solv}(\mathbf{r},\mathbf{r}') = \frac{1}{2}k_B \log\left(\frac{(2\pi)^n}{B}\right) + \frac{3n}{2}k_B \qquad (42)$$

## III. Discussions

Here, we discuss feasibility of calculating $S_{conf}$ and $s_{solv}(\mathbf{r},\mathbf{r}')$ expressed by Eq. (34) and Eq. (42), respectively. The calculation is in fact feasible by means of the statistical mechanics of molecular liquid, or the RISM/3D-RISM theory.[21]

***Feasibility of calculating*** $S_{conf}$ : The expression for $S_{conf}$ involves the second derivative of the free-energy surface with respect to the atomic coordinates of solute, which is defined by Eqs. (13) and (14). The free-energy surface consists of two contributions, the interaction among atoms in the solute, $U(\{\Delta\mathbf{R}\})$, and the solvation free energy, $\Delta\mu(\{\Delta\mathbf{R}\})$. The second derivative of $U(\{\Delta\mathbf{R}\})$ has been calculated rather routinely by means of a molecular mechanics, in the study of the normal mode analysis (NMA) of a biomolecule in *vacuum*.[25]

On the other hand, it is relatively recent that the second derivative of $\Delta\mu(\{\Delta\mathbf{R}\})$ has been calculated by means of the RISM/3D-RISM method. There are two steps in the computation. The first step is to obtain the first derivative, or the force. The method to calculate the force has been proposed by Yoshida and Hirata, which leads,

$$\frac{\partial \Delta\mu}{\partial \Delta\mathbf{R}_\alpha} = \sum_\gamma \rho_\gamma \int d\mathbf{r} \frac{\partial u_\gamma^{uv}(\mathbf{r})}{\partial \Delta\mathbf{R}_\alpha} g_\gamma^{uv}(\mathbf{r}), \qquad (43)$$

where $\rho_\gamma$, $u_\gamma^{uv}(\mathbf{r})$, and $g_\gamma^{uv}(\mathbf{r})$ denote the density of solvent, the interaction between the solute molecule and solvent atom $\gamma$, and the spatial distribution function of solvent atom $\gamma$ at the position **r** around the solute, respectively. [26] The method was applied to the molecular-dynamics simulation by Miyata and Hirata to perform the dynamics of a solute molecule on the free energy surface.[27] The method was further improved by Omelyan and Kovalenko to bring a long simulation such as *protein folding* into a scope of the science. [28] In order to obtain the second derivative, it is necessary to calculate the first derivative of $g_\gamma^{uv}(\mathbf{r})$. The calculation can be performed by applying the procedure originated earlier by Yu and Karplus, in which the second derivative itself is regarded as a variable in the iterative process to find the answer of the RISM/3D-RISM equation. [29,30]

***Feasibility of calcualting*** $s_{solv}(\mathbf{r},\mathbf{r}')$: The calculation is concerned with the matrix $\{\kappa_{ab}^{solv}\}$ defined by Eq. (24), the element of which involves two factors: $J_{ac}(\mathbf{r},\mathbf{r}'')$ and $\chi_{cb}^{(2)}(\mathbf{r}'',\mathbf{r}')$. The calculation of $J_{ac}(\mathbf{r},\mathbf{r}'')$ has been already worked out by Chong and Hirata. [31] On the other hand, the calculation of $\chi_{cb}^{(2)}(\mathbf{r}'',\mathbf{r}')$ is a non-trivial problem, since it is a pair correlation-function of solvent, subject to an inhomogeneous field exerted by a biomolecule. We propose the following superposition-approximation to calculate $\chi_{cb}^{(2)}(\mathbf{r}'',\mathbf{r}')$,

$$\chi_{cb}^{(2)}(\mathbf{r},\mathbf{r}') \approx \langle \rho_c(\mathbf{r}) \rangle \langle \rho_b(\mathbf{r}') \rangle. \tag{44}$$

In the equation, $\langle \rho_c(\mathbf{r}) \rangle$ and $\langle \rho_b(\mathbf{r}') \rangle$ are, respectively, the density distribution functions of atoms *c* and *b* of solvent molecules at the position r and r' around the solute molecule. The quantity can be readily calculated based on the RISM/3D-RISM theory.[32]

## IV. Concluding Remarks and Perspective

Microscopic formula to describe the entropy of a biomolecule in solution are derived based on the Gibbs formula of entropy, and the generalized Langevin theory combined with the RISM/3D-RISM theory. Two types of entropy formula were derived: the conformational entropy and the entropy of solvent around a solute.

The formula of the conformational entropy is closely related to the *inverse* of the *force constant* of the structural fluctuation of a biomolecule in solution, which is the second derivative of the free energy including the potential energy of the solute and the solvation free energy, with respect to the atomic coordinates of the biomolecule. Since the force constant is an inverse of the variance-covariance matrix of the structural fluctuation, the physics implied by the derived formula is in harmony with our intuition; *larger the* *variance, greater the entropy*. A method to calculate the force constant matrix has been already proposed by us, based on the RISM/3D-RISM theory.

The entropy of solvent around a solute molecule is concerned with the inverse of the *force constant*, or the second-derivative, of the free energy of solvent with respect to the density fluctuation of solvent at a pair of positions. The force constant is essentially the inverse of the density pair-correlation function which is a measure of the variance of the density fluctuation. Therefore, the physics is again transparent: *greater the variance of the density fluctuation, larger the entropy*. An approximation to calculate the the density pair-correlation functions in an inhomogeneous environment around a biomolecule is suggested.

In the present paper, we just focused our attention on the entropy of a single solute-molecule in solvent. So, the method *as is* may not be applied to a problem concerning the binding of a ligand molecule by protein. It is because a large change in so-called *external entropies* is expected due to a reduction of the degrees of freedom upon making a complex of the molecules. However, the method developed here may be applied to such a problem by employing the same tactics taken by Chong and Ham in their paper concerning binding of two protein molecules.[3,8] The tactics is to evaluate the external entropy of the bound state by regarding the protein-ligand complex as a "single molecule". The change in the external entropy upon the ligand binding may be calculated by subtracting the independent contributions by host and guest molecules to the external entropy from that by the host-guest complex.

There is another point which was not addressed in the present paper. That is the contribution form multiple conformations of a biomolecule to entropy. The problem may be solved by adapting the same strategy employed in Ref. (29). In the study, the authors have carried out a 3D-RISM/MD simulation of a dipeptide in water to sample the conformational space in order to evaluate the wave-number spectrum of the molecule. They have calculated the Hessian, or the inverse variance-covariance matrix, of the molecule at every few steps and averaged over the trajectory.

# References

[1] Chong, S.-H.; Ham, S. Configurational Entropy of Protein: a Combined Approach Based on Molecular Simulation and Integral-Equation Theory of Liquids. *Chem. Phys. Lett.* **2011**, *504*(46),225–229.

[2] Chong, S.-H.; Ham, S. Conformational Entropy of Intrinsically Disordered Protein. *J. Phys. Chem. B* **2013**, *117*, 5503–5509.

[3] Chong, S.-H.; Ham, S. New Computational Approach for External Entropy in

Protein–Protein Binding. *J. Chem. Theory Comput.* **2016**, *12* (6), 2509–2516.

[4] Lazaridis, T.; Karplus, M. Orientational Correlations and Entropy in Liquid Water. J. Chem. Phys. **1996**, 105, 4294-4316.

[5] Gilson, M. K.; Given, J. A.; Bush, B. L.; McCammon, J. A. The Statistical-Thermodynamics Basis for Computation of Binding Affinity. Biophys. J., **1997**, 72, 1057-1069.

[6] Nguyen, C. N.; Kurtzmann Young, T.; Gilson, M. K. Grid Inhomogeneous Solvation Theory: Hydration Structure and Thermodynamics of the Miniature Recepter Cucurbit[7]Unil. J. Chem. Phys. **2012**, 137, 044101-044118.

[7] Sindhikara, D.; Hirata, F.; Analysis of Biomolecular Solvation Sites by 3D-RISM Theory. J. Phys. Chem. B, **2013**, 117, 6718-6723.

[8] Sugita M.; Kuwano, Izumi.; Higashi, T.; Motoyama, K.; Arima, H.; Hirata, F. Computational Screening of a Functional Cyclodextrin Derivative for Suppressing a Side Effect of Doxorubicin. J. Phys. Chem. B, **2021**, 125, 2308-2316.

[9] Anfinsen, C.B. 1973. Principle that Governs the Folding of Protein Chain. Science, 181: 223-230.

[10] Levinthal, C. Are there pathways protein folding? J. Chim. Phys. 1968, 65, 44-45.

[11] Bryngelson, J.D., Onuchic, J. N., Socci, N. D., Wolynes, P. G., Proteins, **21**, 167, (1995)

[12] Prigogine, I.; Defay, R. *Thermodynamique Chimique (Japanese translation)*. Misuzu publishing Co., 1966.

[13] Rowlinson, J. S. *Liquid and Liquid Mixtures*. Butterworths, 1969, London.

[14] McQuarrie, D. A. *Statistical Mechanics*, Harper and Row, **1976**, New York.

[15] Go, N.; Go, M.; Scheraga, H. A. Molecular Theory of the Helix-Coil Transition in Polyamino Acids, I. Formulation.  Proc. Nat. Acad. Sci. U.S. 1968, 59, 1030-1037.

[16] Go, N.; Scheraga, H. A. J. Chem. Phys. **1969**, 51, 4751-????.

[17] Go, M.; Go, N.; Scheraga, H. A. J. Chem. Phys. **1970**, 52, 2060.

[18] Karplus, M.; Kushick, J. N. Method for estimating the configurational entropy of macromolecules. Macromolecules, 1981, 14, 325-332.

[19] Kim, B.; Hirata, F. Structural Fluctuation of Protein in Water Around Its Native State: New Statistical Mechanics Formulation. *J. Chem. Phys.* **2013**, *138* (5), 054108–054112.

[20] Chandler, D. *Introduction to Modern Statistical Mechanics,* Oxford, 1987. New York.

[21] Hirata, F., Exploring Life Phenomena with Statistical Mechanics of Molecular Liquids, CRC Press, 2020.

[22] Mori, H. Prog. Theor. Phys. 1965, 33, 423.

[23] Kubo, R.；Toda, M.; Hashitsume, N. (eds.) *Statistical Physics II. Non-equilibrium Statistical Mechanics*. Springer, 1992. Berlin.

[24] Hansen, J. P.; McDonald, I. R. "Theory of Simple Liquids". Academic Press. 1986. London.

[25] Go, N.; Noguchi, T.; Nishikawa, Dynamics of a small globular protein in terms of low-frequency vibrational modes. Proc. Natl. Acad. Sci. 1983, 80, 3696-3700.

[26] Yoshida, N.; Hirata, F. A New Method to Determine Electrostatic Potential Around a Macromolecule in Solution from Molecular Wave Functions. J. Comput. Chem., 2006, 27 (4), 453-462.

[27] Miyata, T., Hirata, F. Combination of Molecular Dynamics Method and 3d-RISM Theory for Conformational Sampling of Large Flexible Molecules in Solution. *J. Compt. Chem.* **2007,** 29. 872-882.

[28] Omelyan, I.; Kovalenko, A. J. Chem. Theory. Comput. **2015**, 11, 1875.

[29] Sugita, M., Hirata, F., Realization of the structural fluctuation of biomolecules in solution: Generalized Langevin Mode Analysis, *J. Comp. Chem.*, in press.

[30] Yu, H.-A.; Karplus, M., A thermodynamics Analysis of solvation, *J. Chem.Phys.*,**1988**, 89, 2366.

[31] Chong, S.H.; Hirata, F. Interaction-site-model description of collective excitations in liquid water. I: Theoretical study. J. Chem. Phys. 1999, 111, 3083-3094.

[32] Kovalenko, A.; Hirata, F. Self-consistent description of a metal-water interface by the Kohn-Sham density functional theory and the three-dimensional reference interaction site model. J. Chem. Phys., 1999, 110, 10095-10112.